\documentclass[%
 reprint,
 superscriptaddress,
 amsmath,amssymb,
 aps,
 prl,
]{revtex4-2}

\usepackage{graphicx}% Include figure files
\usepackage{dcolumn}% Align table columns on decimal point
\usepackage{commath}
\usepackage{bm}% bold math
\usepackage{tikz}
\usepackage{comment}
\usepackage[normalem]{ulem}
\usepackage[colorlinks=true,linkcolor=blue]{hyperref}% add hypertext capabilities
\hypersetup{
    colorlinks=true,
    linkcolor=cyan,
    filecolor=magenta,      
    urlcolor=blue,
    citecolor=blue,
}

\usepackage{soul}
\usepackage{lipsum}
\usepackage{xcolor}

\newcommand{\resub}[1]{{\textcolor{red}{\textbf{AM : }}}}

\definecolor{ovdarkgreen}{RGB}{0, 146, 0}

\newcommand{\tikzcircle}[2][ovdarkgreen,fill=ovdarkgreen]{\tikz[baseline=-0.5ex]\draw[#1,radius=#2] (0,0) circle ;}%

\begin{document}

\preprint{APS/123-QED}

\title{Soap Film Drainage Using a Centrifugal Thin Film Balance}

% Force line breaks with \\


\author{Antoine Monier}
\email{antoine.monier95@gmail.com}
\affiliation{Université Côte d'Azur, CNRS, Institut de Physique de Nice (INPHYNI), 06200 Nice, France}

\author{Kévin Gutierrez}%
\affiliation{Université Côte d'Azur, CNRS, Institut de Physique de Nice (INPHYNI), 06200 Nice, France}%

\author{Cyrille Claudet}%
\affiliation{Université Côte d'Azur, CNRS, Institut de Physique de Nice (INPHYNI), 06200 Nice, France}%
 
 \author{Franck~Celestini}
\affiliation{Université Côte d'Azur, CNRS, Institut de Physique de Nice (INPHYNI), 06200 Nice, France}%

\author{Christophe~Brouzet}
\affiliation{Université Côte d'Azur, CNRS, Institut de Physique de Nice (INPHYNI), 06200 Nice, France}%

\author{Christophe~Raufaste}
%\email{christophe.raufaste@univ-cotedazur.fr}
\affiliation{Université Côte d'Azur, CNRS, Institut de Physique de Nice (INPHYNI), 06200 Nice, France} \affiliation{Institut Universitaire de France (IUF), Paris, France}%

\date{\today}% It is always \today, today,

\begin{abstract}
Surface bubbles are an abundant source of aerosols, with important implications for climate processes. In this context, we investigate the stability and thinning dynamics of soap films under effective gravity fields. Experiments are performed using a centrifugal thin-film balance capable of generating accelerations from 0.2 up to 100 times standard gravity, combined with thin-film interferometry to obtain time-resolved thickness maps. Across all experimental conditions, the drainage dynamics are shown to be governed by capillary suction and marginal regeneration—a mechanism in which thick regions of the film are continuously replaced by thin film elements (TFEs) formed at the meniscus. We consistently recover a thickness ratio of 0.8–0.9 between the TFEs and the adjacent film, in agreement with previous observations under standard gravity. The measured thinning rates also follow the predicted scaling laws. We identified that effective gravity has three distinct effects: (i) it induces a strong stretching of the initial film, extending well beyond the linear-elastic regime; (ii) it controls the meniscus size, and thereby the amplitude of the capillary suction and the drainage rate; and (iii) it reveals an inertia-to-viscous transition in the motion of TFEs within the film. These results are supported by theoretical modeling and highlight the robustness of marginal regeneration and capillary-driven drainage under extreme gravity conditions.
\end{abstract}

\maketitle

\twocolumngrid

\section{Introduction}

Surface bubbles are an abundant source of aerosols, whether above a glass of champagne \cite{seon_effervescence_2017} or at the ocean surface \cite{deike_mass_2022}. In the marine environment, the bursting of these bubbles is a primary mechanism for the generation of sea spray aerosols \cite{woodcock_giant_1953, macintyre_flow_1972, wu_evidence_1981, leifer_secondary_2000, bird_daughter_2010}, which play a critical role in atmospheric processes by influencing cloud formation, radiative balance, and the air–sea exchange of climate-relevant gases—making them key players in Earth’s climate system. The stability of surface bubbles is influenced by impurities or surfactants, which enable the thin liquid film to resist perturbations through Marangoni stresses. Nevertheless, the film gradually drains over time due to capillary suction, driven by a pressure difference between the film and the meniscus in contact with the liquid reservoir \cite{lhuissier2012bursting, frostad_dynamic_2016, bhamla_placing_2016,bhamla_interfacial_2017, lin_influence_2018, liu_dynamic_2018, miguet2021marginal}. This drainage results in a net flux from the film to the meniscus, causing progressive thinning and ultimately leading to film rupture as the structure becomes increasingly fragile.

The same drainage process occurs in flat soap films~\cite{gochev2016chronicles}. In this case, the film is planar and connects a rigid frame or a liquid reservoir, terminating at a meniscus with curvature radius $r_{\rm m}$. This curvature induces a pressure deficit of $\gamma/r_{\rm m}$ in the meniscus relative to the flat film, where $\gamma$ is the surface tension. 
Such behavior is well documented in horizontal films, for instance in thin-film balances or equivalent geometries~\cite{sheludko_thin_1967, sonin_role_1993, joye_asymmetric_1994, velev_investigation_1995, cascao_pereira_bike-wheel_2001, yaminsky_stability_2010, chatzigiannakis_thin_2021, andrieux_microfluidic_2021}.    
As in the case of surface bubbles, drainage in flat films is not uniform but is influenced by the instability known as marginal regeneration, a mechanism originally proposed by Mysels \textit{et al.}~\cite{mysels1959soap} and more recently extended by Gros \textit{et al.}~\cite{gros2021marginal}. During this process, the initial film of thickness $h$ feeds into the meniscus, while thinner patches—thin film elements (TFEs) of thickness $h_{\rm TFE}<h$—are intermittently extracted, resulting in a net mass loss from the film. Despite the significance of this process, only a few studies have quantified the thickness contrast between the bulk film and the extracted elements, even though this parameter is essential to predictive models of film stability and drainage \cite{lhuissier2012bursting, gros2021marginal, vigna_flowing_2025}. In vertical films, measurements near the lower horizontal meniscus report a typical ratio $h_{\rm TFE}/h\approx0.8$ \cite{nierstrasz1998marginal, yi_quantitative_2025} where these patches of TFEs detach following  capillary-driven buoyancy \cite{adami2015surface, couder1989hydrodynamics, plateau1873}. 

The nucleation, growth, and detachment of TFEs is critical in the case of surface bubbles, where they invade the entire interface and control the overall thinning dynamics \cite{lhuissier2012bursting, bhamla_placing_2016, frostad_dynamic_2016, miguet2021marginal}. In contrast, in vertical films, TFEs mostly nucleate along the lateral borders, then rise vertically and merge with the film. This leaves the central part of the film relatively smooth, and leads to a characteristic thinning behavior, where the rate is inversely proportional to film width \cite{mysels1959soap, hudales1990marginal, berg2005experimental, seiwert2017velocity, monier_self-similar_2024}, and thickness profiles follow universal self-similar forms \cite{monier_self-similar_2024, auregan_drainage_2025}. In horizontal films, TFEs are also observed, although they remain localized near the meniscus \cite{joye_asymmetric_1994, velev_investigation_1995, khristov_thin_2000, yaminsky_stability_2010, andrieux_microfluidic_2021, gros2021marginal, tregouet2021instability}. According to existing models, the linear flux of liquid exchanged from the film to the meniscus scales as \cite{lhuissier2012bursting, vigna_flowing_2025, cantat_drainage_2026}:
\begin{equation}\label{eq:flux}
q = k \frac{\gamma \, h^{5/2}}{\eta \, r_{\rm m}^{3/2}} ,
\end{equation}
with $\eta$ the liquid viscosity and $k$ a dimensionless factor. While gravity does not appear explicitly in this expression, it influences the meniscus radius $r_{\rm m}$ by setting the curvature difference in surface bubbles \cite{lhuissier2012bursting} or determining meniscus size in planar films \cite{vigna_flowing_2025}. 

\begin{figure*}
    \centering
\includegraphics[width=\linewidth]{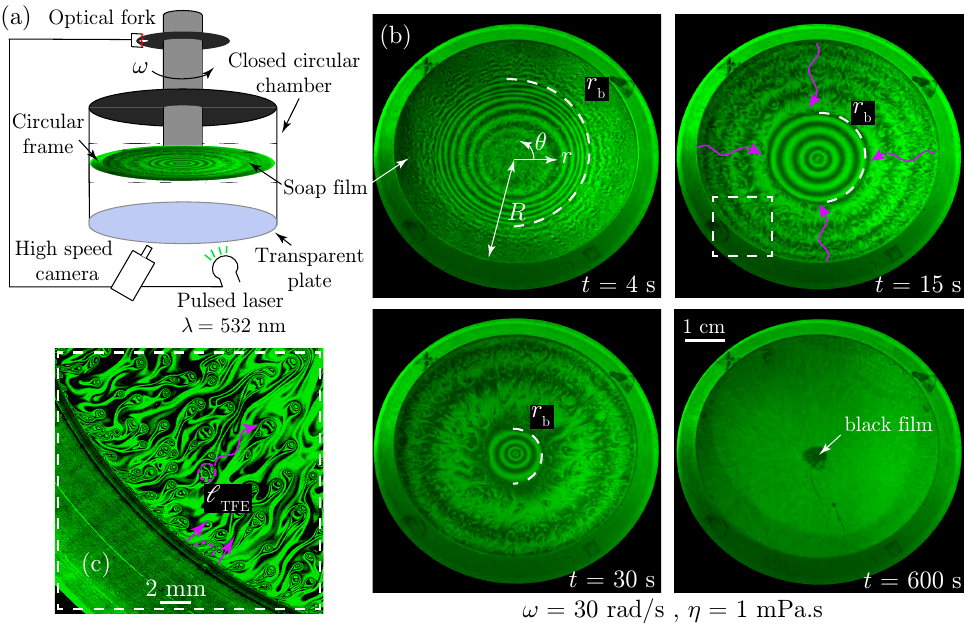}
\caption{
(a) Schematic of the experimental setup. 
(b) Time sequence showing the evolution of a film initially exhibiting two distinct regions (Regime 1). Thin film elements (TFEs) progressively invade the film—first up to the boundary at radius $r_{\rm b}$ (dashed white line), and eventually to the center once $r_{\rm b} = 0$. Note that the central region remains unchanged in thickness but is gradually eroded over time until it vanishes at $r_{\rm b}=0$. The final panel shows the formation and opening of a common black film. The inward migration of TFEs is indicated by wavy magenta arrows.  
(c) Magnified view of the white dashed rectangular area in (b), highlighting the nucleation of TFEs along the edge and their inward migration. $\ell_{\rm TFE}$ denotes the TFE diameter.
}
\label{fig:setup}
\end{figure*}

In this study, we systematically vary the effective gravity and solution viscosity to probe the stability and thinning dynamics of soap films across low and high gravity conditions. 
Our objective is twofold. 
First, we test whether the thickness ratio $h_{\rm TFE}/h$ remains constant or depends on the effective gravity. Second, we examine the validity of Eq.~\eqref{eq:flux} under varying effective gravity, assessing whether marginal regeneration remains the dominant thinning mechanism and whether effective gravity acts only indirectly by setting the meniscus radius $r_{\rm m}$. 
More broadly, this work provides insight into how other body forces—such as magnetic \cite{elias_magnetic_2005, moulton_reverse_2010, back_ferrofilm_2012, lalli_stability_2023} or electric fields \cite{tsekov_streaming_2010, bonhomme_soft_2013, sett_experimental_2016, hussein_sheik_electroosmotic_2017}—may similarly influence drainage dynamics by controlling the meniscus size, and hence the overall liquid exchange between the film and the meniscus.

%In this study, we systematically vary the effective gravity to probe the stability and thinning dynamics of soap films across low and high gravity conditions. We test the robustness of existing models, particularly regarding the apparent universality of the thickness ratio, which remains constant across all gravities and liquid viscosities. A scaling law is established for the exchange flux, confirming that marginal regeneration governs the thinning process. More broadly, this work offers insight into how other body forces—magnetic \cite{elias_magnetic_2005, moulton_reverse_2010, back_ferrofilm_2012, lalli_stability_2023} or electric fields \cite{tsekov_streaming_2010, bonhomme_soft_2013, sett_experimental_2016, hussein_sheik_electroosmotic_2017}—may similarly influence drainage dynamics.

To this end, we developed a centrifugal thin film balance, which allows a soap film to spin about its axis of symmetry. It thereby offers a surface-bubble-like geometry without the complexity of curved interfaces or asymmetric borders typical of vertical films. 
As in surface bubbles, the effective gravity is not spatially homogeneous. In surface bubbles, the relevant body force corresponds to the projection of the gravitational acceleration onto the film surface: its magnitude vanishes along the axis of symmetry and reaches a maximum at the contact between the film and the meniscus. A similar situation occurs in our setup, where the centrifugal acceleration acts as a body force that increases with distance from the axis of rotation. Importantly, its magnitude at the film–meniscus contact can be tuned over more than two orders of magnitude.

%This setup mimics gravity via centrifugal acceleration—achieving variations over more than two orders of magnitude—while suppressing lateral boundaries. 

\section{Methods}

\subsection{Surfactant solutions}

The experiments were conducted using soap films formed from surfactant solutions. These solutions consisted of water–glycerol mixtures containing sodium dodecyl sulfate (SDS) at a concentration of 5.6 g.L$^{-1}$ (2.4~cmc). The water-to-glycerol ratio was systematically varied to tune the bulk viscosity $\eta$ over the range of 1 to 10 mPa.s. Within this composition range, the surface tension remained nearly constant at $35$~ mN.m$^{-1}$~\cite{khan2019effect}. The liquid density was $\rho \gtrsim 1000$~kg$\cdot$m$^{-3}$, with the exact value depending on the water-to-glycerol ratio. Solutions were used within the day of formulation.

\begin{figure*}
\centering
\includegraphics[width=\linewidth]{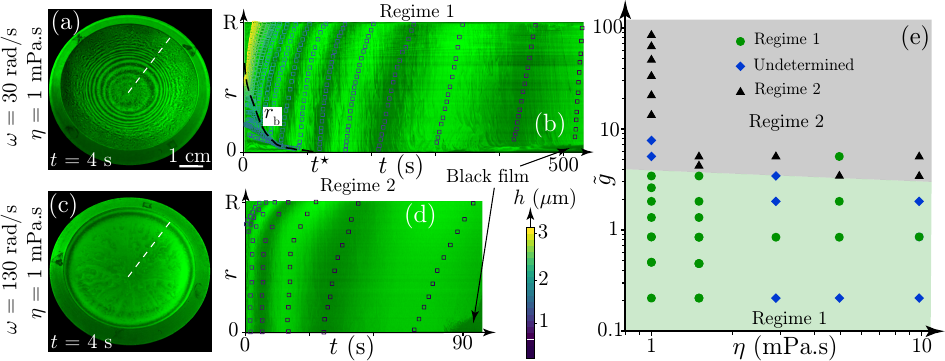}
\caption{
Two drainage Regimes. 
Regime 1: (a) Typical image and (b) space–time diagram constructed along a radius. Isothickness fringes are tracked using color-coded markers that correspond to film thickness. The dashed line indicates $r_{\rm b}(t)$, separating the central smooth region ($0 < r < r_{\rm b}$) from the peripheral zone containing TFEs ($r_{\rm b} < r < R$). After a characteristic time $t^{\star}$, the central region vanishes and TFEs invade the entire film.
Regime 2: (c) Typical image and (d) space–time diagram showing a single region fully populated by TFEs.
(e) Regime diagram as a function of $\tilde{g}$ and $\eta$.  \tikzcircle{2pt} symbols correspond to Regime~1, $\blacktriangle$ to Regime~2, and \textcolor{blue}{$\blacklozenge$} to undetermined cases.}
\label{fig:patch_diagram}
\end{figure*}

\subsection{Centrifugal thin film balance}

The apparatus is depicted in Fig.~\ref{fig:setup}(a). Soap films are formed on a horizontal circular frame with an internal radius $R = 30$~mm by dipping it into the surfactant solution and withdrawing it over approximately one second, at a velocity of about $V \approx 100~\mathrm{mm \, s^{-1}}$. 
The frame is housed within a closed circular chamber, 90~mm in diameter and 40~mm in height, whose lower face is transparent. 
The transparent lower plate serves three purposes: (i) it enables optical access for imaging; (ii) it seals the chamber to maintain a humid environment, thereby suppressing evaporation which reduces the soap film lifetime~\cite{champougny2018influence,miguet2021life,boulogne2022measurement}; and (iii) it entrains the surrounding air in solid-body rotation with the film, ensuring both the film and the air rotate at angular speed~$\omega$. Removing the plate disrupts the film velocity and acceleration conditions.  
There is about 30~s between film formation and the onset of rotation.

At time $t = 0$, the entire system is set into rotation around the vertical axis passing through the center of the film, with an angular speed $\omega$ ranging from 8 to 170~rad.s$^{-1}$ (motor range). 
A transient of approximately 1~s is required for the system to reach rigid-body rotation. %; this timescale is short compared with the thinning dynamics and therefore does not influence the process. 
This rotation induces a centrifugal acceleration~$r \omega^2$, where $r$ is the radial distance from the rotation axis, effectively mimicking a body force whose magnitude increases from the film center to the edge. The resulting acceleration at the film edge can be adjusted between 2 and 840~m·s$^{-2}$, corresponding to approximately 0.2–85 times standard gravity, defining the effective gravity as $\tilde{g} = R \omega^2 / g$.

\subsection{Film imaging and thickness measurements}

The film is illuminated by a pulsed green laser operating at 532~nm (Litron NANO-L-125-15 PIV, 125~mJ per pulse, pulse width 8~ns). The laser is triggered by an optical fork that sends a signal once per rotation, synchronizing the emission of a pulse with each turn. This stroboscopic setup ensures that images are captured at the same angular position on each rotation, effectively freezing any orthoradial motion of the soap film. Images are recorded using a color Nova S12 camera. Both the laser and the camera are positioned at an angle of $70^\circ$ relative to the rotation axis and aligned such that the first reflection from the transparent plate is excluded from the camera’s field of view. Only rays that interfere within the film are collected, resulting in high-contrast interference images, as illustrated in Fig.~\ref{fig:setup}. Once in rotation, the film exhibits concentric black and green interference fringes. At the periphery, the fringes appear rough and irregular due to the presence of TFEs generated along the meniscus. As shown below, these TFEs progressively invade the film, increasingly disturbing the interference pattern. Each fringe corresponds to an isothickness contour, with the thickness difference between a green and a black fringe of 140 nm. To determine the absolute thickness profile, a reference is required; this is taken as the last visible green fringe before the formation of the common black film at the center of the film \cite{gochev2016chronicles,ziapkoff2026white}. At a given time, the film is thinner at its center, with the thickness gradually increasing toward the periphery.

\section{Results and discussion}

\subsection{Typical dynamics}

Upon initiating rotation, the soap film exhibits two distinct drainage Regimes following the transient stage lasting approximately one second.

\textbf{Regime 1} occurs at low effective gravity. It is characterized by the separation of the film into two distinct regions (Fig.~\ref{fig:setup}(b)): a central zone exhibiting smooth, concentric interference fringes, and a peripheral zone where the fringes are partially blurred by the presence of TFEs (Fig.~\ref{fig:setup}(c)). These TFEs nucleate at the edge ($r = R$), migrate inward, and eventually merge with the film at a radius $r = r_{\rm b}$, which defines the boundary between the two regions. As seen in the raw images (Fig.~\ref{fig:setup}(b)) and in the space–time diagram constructed along a radius (Figs.~\ref{fig:patch_diagram}(a) and (b)), $r_{\rm b}$ decreases over time, indicating that the peripheral zone progressively invades the central region until the latter disappears completely at a characteristic time $t^{\star}$. 
However, the fringe dynamics differ between the two zones. In the central region, fringe positions remain nearly stationary over time (visible as ultimately horizontal traces in the space–time diagram), indicating a steady thickness profile. In contrast, in the peripheral zone, the interference fringes are advected outward (corresponding to nearly vertical traces), signaling the progressive thinning of the TFE-containing region, partially blurred. Since the peripheral zone eventually invades the central zone, this mechanism also results in the global thinning of the soap film. Movie1 presents an experiment in Regime 1 in the Supplemental Material.

\textbf{Regime 2} occurs at high effective gravity, with a threshold that shows little dependence on liquid viscosity—typically above 3–5 times standard gravity (Fig.~\ref{fig:patch_diagram}(e)). In this regime, the film no longer exhibits a smooth central zone and is entirely populated with TFEs (Fig.~\ref{fig:patch_diagram}(c)), which still nucleate at the edge but migrate toward the film center. Concurrently, as seen in the space–time diagram (Fig.~\ref{fig:patch_diagram}(d)), the isothickness fringes move outward, indicating global film thinning.
We further observed that in films without glycerol, the effective gravity can be increased up to nearly 100 times standard gravity. However, in glycerol-containing films—where viscosity is higher—the stability decreases markedly above about 5~times standard gravity. Although such films can still be formed, they rupture before reaching the common black film, preventing reliable thickness measurements. An experiment in Regime 2 is shown in Movie2 of the Supplementary Material.

In the following, we denote by $h(r,t)$ the film thickness profile at time $t$. Several characteristic thicknesses are defined: $h_{\rm R}(t) = h(r = R, t)$ corresponds to the film thickness at the frame edge, and $h_{\rm 0}(t) = h(r = 0, t)$ to that at the film center. We also define $h_{\rm b}(t)$ as the thickness at the boundary where TFEs merge with the film. 
Specifically, in Regime~1, $h_{\rm b} = h(r = r_{\rm b}, t)$ for $t < t^{\star}$ while $h_{\rm b} = h_{\rm 0}$ for $t > t^{\star}$. In Regime~2, $h_{\rm b} = h_{\rm 0}$ at all times.

%%%%%%%%%%%%%%%%%%%%%%%%%%%%%%%%%%%%%%%%%%%%%%
\subsection{Initial thickness profiles in Regime 1}

\begin{figure}
\centering
\includegraphics[width=\linewidth]{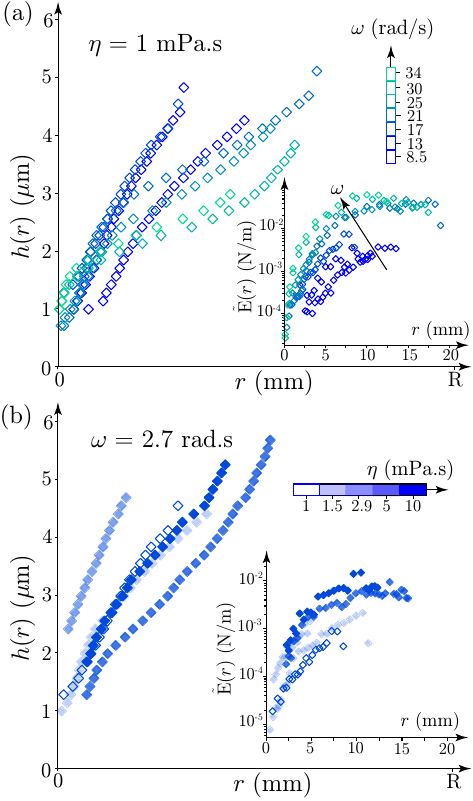}
\caption{
(a) Initial thickness profiles for various rotation speeds at constant viscosity. Inset: corresponding interface elasticity $\tilde{E}(r)$ inferred from the same dataset.
(b) Initial thickness profiles for various viscosities at constant rotation speed.  Inset: corresponding interface elasticity $\tilde{E}(r)$ inferred from the same dataset.}
\label{fig:elasticite}
\end{figure}

In Fig.~\ref{fig:elasticite}, we present the thickness profiles of the central zone in Regime~1. As noted above, these profiles remain essentially unchanged over time, except that the extent of the central zone gradually decreases. We therefore focus on the initial profile, obtained as soon as reliable thickness measurements become available, corresponding to the end of the transient stage.

To investigate the effect of rotation speed, we focus on experiments conducted at the lowest viscosity, for which the dataset is the most extensive. As shown in Fig.~\ref{fig:elasticite}(a), the film thickness increases with radial distance, although not in a strictly regular manner. In all cases, the profile displays an initially convex increase near the center, followed by a concave variation in the middle, and finally a weak concave rise close to the edge. In the convex region near the center, the thickness increases more sharply with higher rotation speeds, indicating that the central part of the film becomes thicker as the rotation rate increases. Conversely, near the periphery, the opposite trend is observed: higher rotation speeds correspond to lower film thicknesses.

When varying the liquid viscosity at a fixed rotation speed, we observe qualitatively similar thickness profiles, although the convex region near the center disappears at the highest viscosities (Fig.~\ref{fig:elasticite}(b)). Aside from this difference, no systematic dependence on viscosity is detected: the profiles are not perfectly superimposed, but the observed variations between experiments do not appear to correlate with viscosity.

We tested several models to understand the initial thickness profiles of the central zone.

Assuming incompressible interfaces~\cite{cantat_drainage_2026}, the internal flow in this geometry reduces to a Poiseuille flow, and the centrifugal force leads to a film of uniform thickness thinning as~\cite{emslie_flow_1958}:
\begin{equation}\label{eq:poiseuille}
h(t) = \frac{h_i}{\sqrt{1+\frac{\rho \omega^2 h_i^2}{12 \eta} (t-t_i)}} ,
\end{equation}
with an initial value $h_{\rm i}$ at time $t_{\rm i}$. 
If we take $t_{\rm i} = 0$ and $h_{\rm i} = h_0$, the film thickness prior to centrifugation, we can estimate the thinning of the central zone during the transient stage associated with  the internal flow. The thickness $h_0$ is estimated using Frankel’s law~\cite{champougny2015surfactant}, $h_{\rm 0} \simeq 2 \kappa^{-1} \left( \frac{\eta V}{\gamma_{\rm 0}} \right)^{2/3}$, with $\kappa^{-1} \approx 2~\mathrm{mm}$ the capillary length and $V \approx 100~\mathrm{mm \, s^{-1}}$ characteristic withdrawal speed to form the film. This yields $h_{\rm 0} \approx 80~\mu\mathrm{m}$ for the lowest viscosity and $400~\mu\mathrm{m}$ for the highest. 
The central region is observed only for $\omega \lesssim 40$~rad·s$^{-1}$, regardless of viscosity. Taking this value as an upper bound in Eq.~\eqref{eq:poiseuille} and using $t = 1$~s leads to a predicted thickness reduction of 25–40\%, depending on viscosity, by the end of the transient. These values are overestimates, since Eq.~\eqref{eq:poiseuille} assumes steady rotation, whereas the system does not reach its maximum rotation speed during the transient stage. At most, we therefore expect $h_0$ to decrease to about $60~\mu$m, which is still more than an order of magnitude larger than the few-micron thicknesses observed experimentally in the central zone. This indicates that internal flow may contribute weakly during the transient, but is not the dominant thinning mechanism at this stage.

%Using typical values $\rho = 1000$~kg·m$^{-3}$, $\eta = 1$~mPa·s, and $\omega = 100$~rad·s$^{-1}$, the film thickness reaches about 1~µm after approximately $10^3$~s. This timescale is much longer than the duration over which the central region is observed, indicating that internal flow can be neglected—consistent with the fact that the profiles do not show a strong dependence on liquid viscosity. 

As a consequence, film stretching due to interfacial elasticity appears to be the most plausible mechanism to describe the thickness profiles in the central region~\cite{cantat_drainage_2026}. The observed profiles are therefore compared with a theoretical model based on the following assumptions. Since these profiles are not yet populated with TFEs, we assume they result from the stretching of an initially uniform film formed prior to centrifugation. Following the approach of Pasquet \textit{et al.}~\cite{pasquet_thickness_2023} for vertical films under standard gravity, we assume that the film has an initial uniform thickness $h_{\rm 0}$ and surface tension $\gamma_{\rm 0}$ before rotation begins. Once the film is set into rotation, it is stretched by centrifugal forces, which are balanced by surface tension gradients~\cite{couder1989hydrodynamics, de2001young}, leading to 
\begin{equation}\label{eq:gamma_gradient}
\frac{\partial \gamma}{\partial r} = -h \rho r \omega^2/2 .
\end{equation}
Because the initial surfactant concentration is above the critical micelle concentration (cmc), the film elasticity is assumed to be primarily associated with the presence of surface-active impurities \cite{cantat_drainage_2026}. As the film stretches and thins, the local interfacial concentration of these impurities decreases. 
The surface tension can be approximated by $\gamma = \gamma_{\rm 0} + E\frac{\varepsilon}{1+\varepsilon}$, where $E$ denotes the interfacial elasticity and $\varepsilon=(h_0-h)/h$ the interface extension~\cite{poryles2022non, pasquet_thickness_2023}. 
Combining these relations yields the following differential equation for the thickness $h(r)$:
\begin{equation}\label{eq:prediction_diff_eq_h}
\frac{d h(r)}{d r}   = \frac{\rho h_{\rm 0} \omega^2}{2 E} r h(r),
\end{equation}
and the following expression for the thickness profile:
\begin{equation}\label{eq:prediction_h_stretching}
h(r) = h(r=0) e^{(r/r_{\rm 0})^2}  ,
\end{equation}
where $h(r=0)$ is the film thickness at the center and $r_{\rm 0} = \sqrt{\frac{4 E}{\rho \omega^2 h_{\rm 0}}}$ is the characteristic length scale governing the radial thickness variation.

Eq.~\eqref{eq:prediction_h_stretching} predicts a sharp increase in thickness due to the combined exponential and quadratic dependencies. This prediction does not capture the experimentally observed irregular increase of $h(r)$ with $r$ (Fig.~\ref{fig:elasticite}). Even though the theoretical approach assumes a constant interfacial elasticity, we can use Eq.~\eqref{eq:prediction_diff_eq_h} to infer an apparent local elasticity, $\tilde{E}(r) = \frac{\rho h_{\rm 0} \omega^2}{2} \, \frac{r}{\frac{d}{dr}\left( \ln h(r)\right)}$. 
%\textcolor{blue}{\sout{To evaluate this expression, we estimate the initial thickness before rotation using Frankel’s law~\cite{champougny2015surfactant}, $h_{\rm 0} \simeq 2 \kappa^{-1} \left( \frac{\eta V}{\gamma_{\rm 0}} \right)^{2/3}$, with $\kappa^{-1} \approx 2~\mathrm{mm}$ the capillary length and $V \approx 100~\mathrm{mm \, s^{-1}}$ the characteristic pulling speed used to form the film. This yields $h_{\rm 0} \approx 80~\mu\mathrm{m}$.}} 

In the inset of Fig.~\ref{fig:elasticite}(a), $\tilde{E}(r)$ is shown as a function of rotation speed for the solution without glycerol. In all cases, $\tilde{E}$ varies significantly across the radius—by up to three orders of magnitude—and the larger the rotation speed, the larger the variation. At the inner radius, values start around $10^{-5}$–$10^{-4}$~N·m${}^{-1}$ and increase to $3 \times 10^{-3}$–$3 \times 10^{-2}$~N·m${}^{-1}$, eventually reaching a plateau at the highest rotation speeds. These values can be compared to $10^{-3}$~N·m${}^{-1}$, the typical value reported by Poryles \textit{et al.}~\cite{poryles2022non} for the same surfactant solution and concentration (2.4 times above the cmc), associated with the presence of dodecanol molecules acting as insoluble surface-active species at the interface. While this value lies near the middle of the range observed at a given rotation speed, our data cannot be fully explained by this description of interfacial elasticity. One possible explanation is the very large interface extension observed in our experiments, estimated as $(h_0 - h(r=0))/h(r=0)$, with values around 80—well beyond unity—leading to strongly nonlinear behavior and combined effects of both dodecanol and SDS molecules on the surface tension, even above the cmc, due to exchanges between the bulk and the interfaces~\cite{poryles2022non}. As discussed previously, interface stretching may also couple with internal flow at the highest rotation speeds. A similar trend is observed  when varying the viscosity at constant rotation speed (inset of Fig.~\ref{fig:elasticite}(b)). 
These observations call for an improved description of film elasticity to account for interface stretching and film thinning during the transient stage. 

This analysis also highlights the significant thinning that occurs during the transient stage, until the body force reaches its steady-state value, during which the film thickness decreases from 80–400~$\mu$m to a few microns within typically 1~s. In what follows, we focus on the thinning dynamics after this transient stage, dominated by the dynamics of the TFEs.

%Thinning at the onset of rotation—prior to the development of marginal regeneration—cannot be explained by internal flow or by film stretching within the framework of current models. This calls for an improved description of film elasticity to account for this stage. However, in the following, a detailed characterization of the central zone thickness in Regime~1 is not required to understand the mechanisms of marginal regeneration, which govern film thinning at later times. 

%%%%%%%%%%%%%%%%%%%%%%%%%%%%%%%%%%%%%%%%%%%%%%
\begin{figure}
    \centering
    \includegraphics[width=\linewidth]{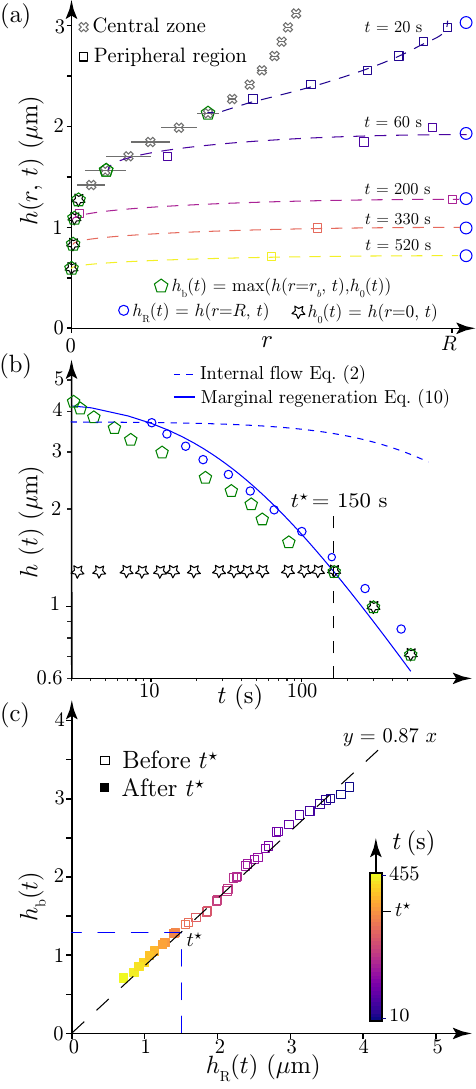}
\caption{
Thickness measurements for the experiment shown in Fig.~\ref{fig:setup}.  
(a) Profiles at different times (colors). Squares denote values measured in the peripheral region (dashed lines are guides to the eye), while crosses indicate the mean thickness profile of the central zone, with horizontal bars representing the standard deviation of the position. The symbols corresponding to $h_{\rm R}$, $h_0$, and $h_{\rm b}$ are defined in the panel. 
(b) Temporal evolution of $h_{\rm R}(t)$, $h_{\rm 0}(t)$, and $h_{\rm b}(t)$ on a log–log scale. Data are shown for every other interference fringe for clarity. The time $t^\star$ marks the full erosion of the central zone, defined by $h_{\rm b}(t^\star)=h_{\rm 0}(t^\star)$. 
Same symbols as in (a). (c) $h_{\rm b}(t)$ as a function of $h_{\rm R}(t)$, with open and closed symbols corresponding to times before and after $t^\star$, respectively. The dashed line shows a proportionality of 0.87.}
\label{fig:time_thickness}
\end{figure}

\subsection{Thickness measurements}

\begin{figure}
    \centering
\includegraphics[width=\linewidth]{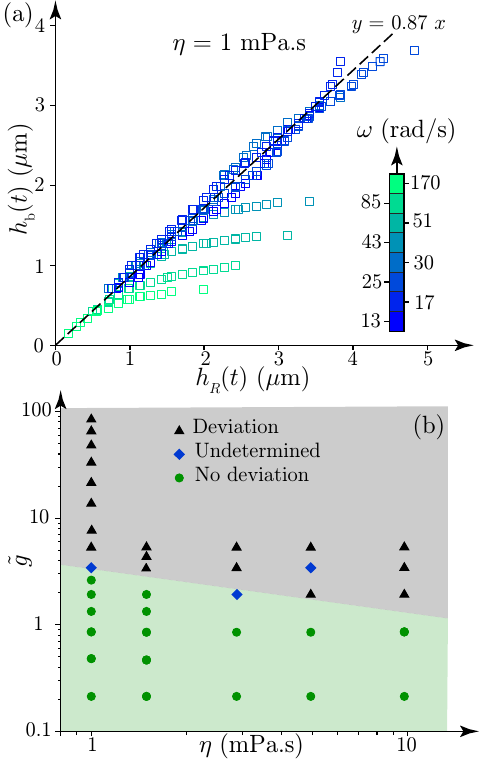}
\caption{
(a) $h_{\rm b}$ as a function of $h_{\rm R}$ for various times and rotation speeds, showing an initial deviation from proportionality at the highest rotation speeds. The dashed line indicates a proportional relationship with a coefficient of 0.87.
(b) Deviation diagram as a function of $\tilde{g}$ and $\eta$. \tikzcircle{2pt} symbols correspond to cases with no deviation from proportionality, $\blacktriangle$ to cases with deviation, and \textcolor{blue}{$\blacklozenge$} to undetermined cases.}
    \label{fig:hc}
\end{figure}

In Fig.~\ref{fig:time_thickness}(a), we present typical time evolutions of full thickness profiles $h(r, t)$ in Regime 1, illustrating the coexistence of the two regions until the outer region fully invades the center. 

As observed in the space–time diagram of Fig.~\ref{fig:patch_diagram}(b), the positions of the isothickness fringes in the central zone remain nearly stationary over time, indicating a steady thickness profile. This behavior is reflected by the crosses in Fig.~\ref{fig:time_thickness}(a), which represent time-averaged values, with error bars corresponding to the temporal standard deviations. This observation indicates that internal flow described by Eq.~\eqref{eq:poiseuille} can be neglected in the central zone. Using typical values $\rho = 1000$~kg·m$^{-3}$, $\eta = 1$~mPa·s, $\omega = 40$~rad·s$^{-1}$, and $h_i = 5~\mu$m, the time required for the film thickness to decrease by a factor of two is of order $36\eta/(\rho \omega^2 h_i^2) \sim 900$~s. This timescale is consistently much longer than the lifetime of the central region, $t^\star$, confirming that internal flow is negligible over this duration.

In Fig.~\ref{fig:time_thickness}(b), we plot $h_{\rm R}$, $h_{\rm 0}$ and $h_{\rm b}$ as functions of time for the same dataset. 
Since the thickness profile in the central region remains steady, $h_{\rm 0}$ stays constant up to $t^{\star}$, the time at which the central zone disappears. 
We observe that both $h_{\rm R}$ and $h_{\rm b}$ decrease over time. Moreover, a proportionality relation is verified between these two quantities: as shown in Fig.~\ref{fig:time_thickness}(c), plotting $h_{\rm b}$ against $h_{\rm R}$ reveals a proportionality with a coefficient of approximately 0.87, which holds even at early times.

In Fig.~\ref{fig:hc}(a), we plot these two quantities for the solution with viscosity $\eta = 1$~mPa·s and for several rotation speeds, thus covering both Regimes~1 and~2. The proportionality relation between $h_{\rm b}$ and $h_{\rm R}$ holds at the lowest rotation speeds, with a proportionality coefficient of $0.87 \pm 0.01$, obtained from a best fit over all corresponding data. At higher rotation speeds, a deviation from this proportionality appears at early times, when the film is relatively thick. We verified that these behaviors are consistent across all tested viscosities. We define~$h_{\rm c}$ as the critical thickness below which the proportionality is recovered, and we find that $h_{\rm c}$ decreases with increasing rotation speed. In Fig.~\ref{fig:hc}(b), we present an occurrence diagram distinguishing experiments that exhibit early-time deviations from proportionality. A comparison with the Regime diagram in Fig.~\ref{fig:patch_diagram}(c) reveals a strong correlation: full proportionality is observed primarily in Regime~1, while partial proportionality systematically occurs in Regime~2, as well as in a few high-viscosity cases at the highest rotation speeds within Regime~1.

These observations lead to two key remarks:

First, when proportionality is observed, the coefficient of 0.87 is consistent with the value reported by Nierstrasz \textit{et al.}~\cite{nierstrasz1998marginal} for vertical films under standard gravity. This supports the interpretation that TFEs nucleate along the frame edge, detach, and migrate radially inward under effective buoyancy, ultimately settling where the film thickness matches that of the TFEs (see the nucleation along the frame edge in Fig.~\ref{fig:setup}(c), as well as the inward migration indicated by the arrows in panels (b) and (c) of the same figure). 
This scenario assumes that the TFE migration time is short compared to the overall thinning time, which is always the case in our observations.  
Our findings validate this mechanism even under centrifugal acceleration up to almost 100 times standard gravity. The gravity-independence of the 0.87 ratio reinforces the scenario proposed by Gros \textit{et al.}~\cite{gros2021marginal}, which attributes the selection mechanism to capillary suction and Marangoni stresses, rather than gravitational forces. 

\begin{figure}
    \centering
\includegraphics[width=\linewidth]{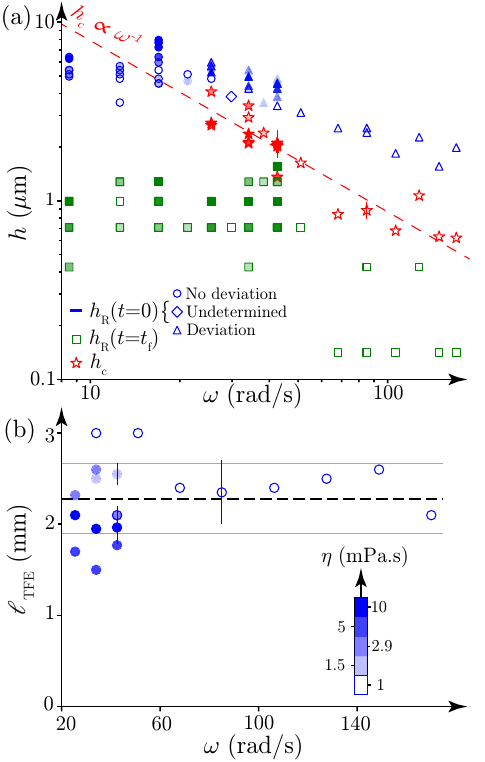}
\caption{
(a) $h_{\rm R}(t=0)$, $h_{\rm c}$ (when a deviation is observed), and $h_{\rm R}(t=t_{\rm f})$ as functions of $\omega$, where $t_{\rm f}$ denotes the final measurement time. Data from all experiments and viscosities are shown with opacity encoding.
(b) $\ell_{\rm TFE}$ as a function of $\omega$, showing that this quantity remains approximately constant. Measurements are taken when $h_{\rm R} = h_{\rm c}$, while TFEs are located at the film center. Data include all experiments in which a deviation from proportionality is observed.}
    \label{fig:hc_leff}
\end{figure}

Second, measurements of the critical thickness $h_{\rm c}$ indicate a weak dependence on liquid viscosity, but a clear decrease with increasing rotation speed (see Fig.~\ref{fig:hc_leff}(a)). 
We propose that this behavior results from an inertia-to-viscous transition in the motion of the TFEs. 
Specifically, assuming that the proportionality between the TFE thickness and $h_{\rm R}$ still holds, inertia prevents TFEs from stopping precisely where their thickness matches that of the surrounding film. As a result, $h_{\rm b}$ no longer corresponds to the local TFE thickness, explaining the observed deviation from proportionality. 
Considering a TFE patch of diameter $\ell_{\rm TFE}$ and thickness $h_{\rm TFE}$ (see Fig.~\ref{fig:setup}(c)), Newton's second law applied along the radial direction gives:
\begin{equation}\label{eq:tfe_dynamics}
\frac{1+\alpha}{4} \rho h_{\rm TFE} \pi \ell_{\rm{TFE}}^2 \frac{d^2 r}{dt^2} + \eta^{2D} \frac{dr}{dt} + \rho \frac{\ell_{\rm{TFE}}^2}{4} r \omega^2 (h(r) - h_{\rm TFE}) = 0 ,
\end{equation}
which is the equation of an oscillator around the position where $h = h_{\rm TFE}$ as expressed in the third term. The first term corresponds to the inertial force, with a patch mass of $\rho h_{\rm TFE} \pi \ell_{\rm{TFE}}^2/4$ and added mass factor $\alpha$. The second term is a damping force with contributions from surface viscosity $\eta_{\rm s}$, bulk viscosity $\eta$, and air friction (with viscosity $\eta_{\rm{air}}$), such that the total two-dimensional viscosity is $\eta^{2D} = 2 \eta_{\rm s} + \eta h + k_{\rm{air}} \eta_{\rm{air}} \ell_{\rm{TFE}}$, where $k_{\rm{air}}$ is a geometric prefactor \cite{louyer_sliding_2025}. 
When damping dominates, TFEs directly reach their equilibrium position and $h_{\rm TFE}=h_{\rm b}$. This is consistent with observations in Regime 1 and Regime 2 when $h_{\rm R} < h_{\rm c}$. In contrast, during the early stages of Regime 2 (when $h_{\rm R} > h_{\rm c}$), inertia dominates, and TFEs overshoot their equilibrium and reach the film center.

At the transition, we expect all terms in Eq.~\eqref{eq:tfe_dynamics} to be of comparable magnitude. Taking a perturbation of amplitude $a$ and time scale $\tau$, the three terms scale as $\rho h_{\rm c} \ell_{\rm{TFE}}^2 a / \tau^2$, $\eta^{2D} a / \tau$, and $\rho h_{\rm c} \ell_{\rm{TFE}}^2 a \omega^2$, respectively. Equating these scales gives $\tau \sim 1/\omega$ and:
\begin{equation}\label{eq:hc_eq1}
h_{\rm c} \sim \frac{\eta^{2D}}{\rho \ell_{\rm{TFE}}^2 \omega} .
\end{equation}
%In experiments, we found that $\ell_{\rm{TFE}}$ can be considered approximately constant (around 2~mm), both at the boundary between the two regions in Regime~1 and at the film center in Regime~2, when $h_{\rm R}$ reaches $h_{\rm c}$ (Fig.~\ref{fig:hc_leff}(b)). This value appears to be independent of $\eta$ and $\omega$, allowing $\ell_{\rm TFE}$ to be treated as constant in Eq.~\eqref{eq:hc_eq1}. A detailed space–time analysis of $\ell_{\rm TFE}$ is beyond the scope of the present study.
The diameter of TFE $\ell_{\rm{TFE}}$ has been measured in all experiments, both at the boundary between the two regions in Regime~1 and at the film center in Regime~2 (when $h_{\rm R}$ reaches $h_{\rm c}$). Figure~\ref{fig:hc_leff}(b) shows that $\ell_{\rm{TFE}}$ appears to be independent of $\omega$, and weakly dependent of $\eta$ only. In Eq.~\eqref{eq:hc_eq1}, we therefore consider $\ell_{\rm{TFE}}$ as a constant, approximately equal to $2$~mm. A detailed space–time analysis of $\ell_{\rm TFE}$ is beyond the scope of the present study. \\

Eq.~\eqref{eq:hc_eq1} therefore predicts that $h_{\rm c}$ should scale as $1/\omega$, in good agreement with our measurements (Fig.~\ref{fig:hc_leff}(a)).
Furthermore, this framework explains the absence of deviations from proportionality at the lowest rotation speeds. Indeed, comparing the trend obtained for $h_{\rm c}$ with the initial edge thickness $h_{\rm R}(t=0)$ at the onset of measurements, we note that $h_{\rm R}(t=0)$ decreases with increasing $\omega$ for all viscosities, but more weakly than the $\omega^{-1}$ dependence of $h_{\rm c}$ (Fig.~\ref{fig:hc_leff}(a)).
As a consequence, deviations from proportionality occur only when $h_{\rm R}(t=0) > h_{\rm c}$, which does not happen at the lowest rotation speeds. 

Quantitatively, within $\eta^{2D}$, the surface viscosity $\eta_{\rm s}$ is typically smaller than $10^{-8}$~Pa$\cdot$s$\cdot$m \cite{zell2014surface, louyer_sliding_2025}, the bulk viscosity contribution $\eta h$ is around $10^{-9}$~Pa$\cdot$s$\cdot$m, and the air friction term $\eta_{\rm{air}} \ell_{\rm{TFE}}$ is approximately $10^{-8}$~Pa$\cdot$s$\cdot$m. This suggests that air friction is the dominant damping mechanism, consistent with previous observations \cite{lenavetier2024line, louyer_sliding_2025}, and explains the insensitivity of $h_{\rm c}$ to liquid viscosity. 
For a typical rotation speed of $\omega = 100$~rad$\cdot$s$^{-1}$, this model predicts $h_{\rm c} \sim 0.1~\mu$m, in qualitative agreement with our experimental observations ($h_{\rm{c}} \lesssim 1~\mu$m, see Fig.~\ref{fig:hc_leff}(a)), supporting the hypothesis of an inertia-to-viscous transition occurring at $h_{\rm c}$. The discrepancy remains below one order of magnitude, which is satisfactory for a scaling estimate.

\subsection{Thinning dynamics}

\begin{figure}
    \centering
\includegraphics[width=\linewidth]{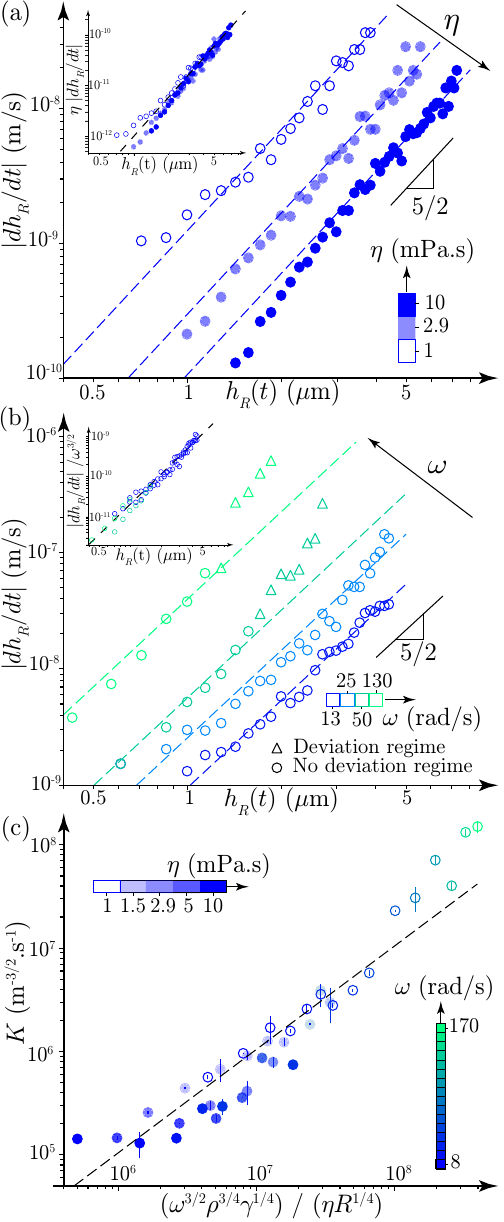}
\caption{
(a) $\lvert \textrm{d}h_{\rm R}/\textrm{d}t \rvert$ as a function of $h_{\rm R}$ at constant $\omega$ and varying $\eta$, in opacity. Insert: $\eta \, \lvert \textrm{d}h_{\rm R}/\textrm{d}t \rvert$ as a function of $h_{\rm R}$ for the same data. 
(b) $\lvert \textrm{d}h_{\rm R}/\textrm{d}t \rvert$ as a function of $h_{\rm R}$ at constant $\eta$ and various $\omega$, in colors. Insert: $\lvert \textrm{d}h_{\rm R}/\textrm{d}t \rvert/ \omega^{3/2}$ as a function of $h_{\rm R}$ for the same data. 
(c) $K$ as a function of the prediction for all data, with same color-coding as (a,b).}
\label{fig:dh_dt}
\end{figure}

The time evolution of the various film thicknesses shown in Fig.~\ref{fig:time_thickness}(b) allows for comparison across experiments. We assume that the mean film thickness scales with $h_{R}$, as supported by the observed proportionality relation, except in the central zone, whose area is typically small compared to the TFE-populated region. 

As illustrated by the example in Fig.~\ref{fig:time_thickness}(b), internal flow described by Eq.~\eqref{eq:poiseuille}, in blue dashed line, is too slow to account for the experimentally observed thinning dynamics, indicating that marginal regeneration is the dominant thinning mechanism in this case. This conclusion remains valid even at the highest effective gravities, as detailed in the Discussion and supported by the analysis that follows.

Assuming thinning driven by marginal regeneration, the expression for the linear flux~$q$, as given in Eq.~\eqref{eq:flux}—previously validated under standard gravity~\cite{lhuissier2012bursting, vigna_flowing_2025, cantat_drainage_2026}—suggests a correlation between the rate of thinning $\textrm{d} h_{\rm R}/\textrm{d}t$ and the film thickness $h_{\rm R}$. Approximating the film volume as $\pi R^2 h_{\rm R}$ and the liquid loss rate as $2 \pi R q$, we expect, following~\cite{lhuissier2012bursting, vigna_flowing_2025, cantat_drainage_2026}:
\begin{equation}\label{eq:relation_dhdt_h}
\frac{\textrm{d} h_{\rm R}}{\textrm{d} t} \propto - \frac{\gamma}{\eta} \frac{h_{\rm R}^{5/2}}{R r_{\rm m}^{3/2}}.
\end{equation}

In Fig.~\ref{fig:dh_dt}(a), we plot $\lvert \mathrm{d}h_{\rm R}/\mathrm{d}t \rvert$ as a function of $h_{\rm R}$ for a fixed rotation speed and various liquid viscosities. For all cases, we observe a power-law dependence with an exponent close to $5/2$, and thinning is slower for higher viscosities, as expected in Eq.~\eqref{eq:relation_dhdt_h}. When plotting $\eta\,  \lvert \mathrm{d}h_{\rm R}/\mathrm{d}t \rvert$ as a function of $h_{\rm R}$ (see inset of Fig.~\ref{fig:dh_dt}(a)), all data collapse onto a single curve, confirming the validity of Eq.~\eqref{eq:relation_dhdt_h}. 

In Fig.~\ref{fig:dh_dt}(b), we explore the influence of rotation speed by plotting $\lvert \mathrm{d}h_{\rm R}/\mathrm{d}t \rvert$ versus $h_{\rm R}$ at fixed viscosity. Each dataset again follows a power law with an exponent close to $5/2$. Deviations appear when $h_{\rm R} > h_{\rm c}$, as shown by the triangles markers in Fig.~\ref{fig:dh_dt}(c) where the proportionality between $h_{\rm R}$ and $h_{\rm b}$ no longer holds; these data are excluded from the analysis that follows. As expected, higher rotation speeds lead to faster thinning. A power-law fit yields $\lvert \mathrm{d}h_{\rm R}/\mathrm{d}t \rvert \propto \omega^{1.31 \pm 0.33}$. Although $\omega$ does not appear explicitly in Eq.~\eqref{eq:relation_dhdt_h}, it influences the meniscus radius $r_{\rm m}$, which we assume to scale with the effective capillary length, $r_{\rm m} \propto \sqrt{\gamma / (\rho R \omega^2)}$. Substituting this relation in Eq.~\eqref{eq:relation_dhdt_h} gives the thinning law:  
\begin{equation}\label{eq:thinning_MR}
%\left\lvert \frac{\textrm{d} h_{\rm R}}{\textrm{d} t} \right\rvert \propto
\frac{\textrm{d} h_{\rm R}}{\textrm{d} t} \propto -\frac{\omega^{3/2} \rho^{3/4} \gamma^{1/4}}{\eta \, R^{1/4}} \, h_R^{5/2} .
\end{equation}
Noting $K \propto {\omega^{3/2} \rho^{3/4} \gamma^{1/4}}/({\eta \, R^{1/4}})$ 
the proportionality constant between $\lvert\textrm{d} h_R/\textrm{d}t \rvert$ and $h_R^{5/2}$, %\add{where $k$ is a dimensionless constant without unity, }
the predicted dependence $K \propto \omega^{3/2}$ agrees well with the empirical exponent $1.31 \pm 0.33$. When $\omega^{-3/2} \, \lvert \mathrm{d}h_{\rm R}/\mathrm{d}t \rvert$ is plotted against $h_{\rm R}$ for the data in Fig.~\ref{fig:dh_dt}(b), all curves collapse onto a single master curve (see inset), confirming the predicted $\omega$ scaling.

Overall, the expression for $K$ is validated over 3 orders of magnitude in Fig.~\ref{fig:dh_dt}(c), where the effects of both viscosity and rotation speed were independently tested.
These results confirm thinning driven by marginal regeneration and that  Eq.~\eqref{eq:flux} remains valid even under high-gravity conditions. %\add{A fit of the data in Fig.~\ref{fig:dh_dt}(c) leads to a dimensionless constant $k\approx0.1$. 
By integrating Eq.~\eqref{eq:thinning_MR} over time, we obtain a formula analogous to Eq.~\eqref{eq:poiseuille} for internal flow, but describing thinning driven by marginal regeneration:
\begin{equation}\label{eq:thinning_MR_h}
h_{\rm R}(t)=\frac{h_i}{\left(1+\frac{3}{2} K h_i^{3/2}(t-t_i)\right)^{2/3}},
\end{equation}
where $h_i$ is the initial thickness at time $t_i$. Testing this expression on the data from Fig.~\ref{fig:time_thickness}(b) using the corresponding value of $K$ from Fig.~\ref{fig:dh_dt}(c) shows good agreement. Furthermore, Eq.~\eqref{eq:thinning_MR_h} exhibits the asymptotic scaling $h_{\rm R} \propto t^{-2/3}$, consistent with previous studies under standard gravity~\cite{lhuissier2012bursting, miguet2021marginal}.
%They are also consistent with the asymptotic behavior $h_{\rm R} \propto t^{-2/3}$ observed at long times (Fig.~\ref{fig:time_thickness}~(b)), in agreement with previous studies under standard gravity~\cite{lhuissier2012bursting, miguet2021marginal}. \chris{ici je mettrai carrément la formule de $h_R$ que l'on peut obtenir à partir de \eqref{eq:thinning_MR} et qu'Antoine a utilisé dans la fig.~4(b):}

\subsection{Discussion}

In this work, we investigated the influence of effective gravity—ranging from 0.2 to 100 times standard gravity—on the thinning dynamics of soap films, while also varying the viscosity of the surfactant solution.

At early stages, in the absence of marginal regeneration, centrifugal acceleration likely induces an initial stretching of the film rather than thinning through internal flow, although existing interfacial elasticity models fail to reproduce the observed profiles. This stretching occurs only in the central region and at low effective gravity, in Regime~1, before erosion by marginal regeneration begins. 
In Regime~2, TFEs populate the entire film as soon as imaging begins, suggesting that either the initial erosion of the central zone is too fast to be captured, or that the film is sufficiently flat for TFEs to migrate directly toward the center. 
Interestingly, films remain very stable, particularly in the absence of glycerol. Assuming an average film thickness of a few micrometers, the surface tension difference between the center and the frame edge can be estimated from Eq.~\eqref{eq:gamma_gradient} as $\Delta \gamma = \rho h R^2 \omega^2 / 4 \sim 7$~mN·m$^{-1}$, smaller than but comparable to the surface tension of the surfactant solution. This difference may explain why the system approaches the limit of film stability reported in previous studies \cite{de2001young}.

The presence or absence of a central zone remains one of the open questions of this study. The transition between Regimes~1 and~2 occurs at effective gravities of approximately 3–5 times standard gravity, with only a weak dependence on viscosity. Investigating the transient stage during which the film is brought into solid-body rotation is therefore crucial for understanding the early-stage dynamics and, in particular, the fate of the central zone, as well as how marginal regeneration operates during this period—potentially coupled with interface elasticity and, to a lesser extent, internal flow. Future work could address this issue by controlling the duration of this transient stage—about 1~s in the present study—for instance by progressively ramping up the rotation speed.

After the transient stage, once the system has reached solid-body rotation, thinning is unambiguously associated with marginal regeneration. Indeed, comparing the thinning rate due to marginal regeneration, given by Eq.~\eqref{eq:thinning_MR}, with that due to internal flow, $\lvert \mathrm{d}h_R/\mathrm{d}t \rvert \approx \rho \omega^2 h_R^3 / \eta$, obtained from Eq.~\eqref{eq:poiseuille}, shows that marginal regeneration dominates as long as
$h_R \ll \sqrt{\gamma/(\rho R \omega^2)}$. The length scale on the right-hand side corresponds to the effective capillary length that already sets the meniscus radius $r_{\rm m}$. In our parameter range, it typically exceeds $200~\mu$m, and is therefore much larger than the values of $h_R$ once the system is in solid-body rotation. 

Across all Regimes and viscosities, thinning is consistently associated with the nucleation of TFEs in the vicinity of the meniscus. These elements are thinner than the surrounding film, resulting in a net mass loss as they migrate inward. On average, we find a proportionality factor of $0.87 \pm 0.01$ between the TFE thickness and the local film thickness, independent of the effective gravity, viscosity, or thinning Regime. This value is consistent with previous measurements under standard gravity~\cite{nierstrasz1998marginal, yi_quantitative_2025} and slightly higher than the theoretical maximum of approximately 0.7 predicted by current models of marginal regeneration~\cite{mysels1959soap, gros2021marginal, cantat_drainage_2026}. 
The dynamics of these thin film elements depend on the effective gravity. In Regime 1, at low effective gravity, the film exhibits a strong thickness gradient, and TFEs stop at the boundary where their thickness equals the local film thickness. The central region is progressively eroded, producing a gradual reduction of the thickness gradient with a factor of 0.87 from the film center to the edge. In Regime~2, where TFEs populate the film from the start, the same proportionality factor of 0.87 is observed directly between the center and the boundary. In both regimes, these observations are modulated by inertial effects when $h_R$ is initially sufficiently thick.

Although effective gravity modifies the overall thinning patterns—leading to two distinct regimes and an inertia-to-viscous transition—it does not alter the fundamental expression for the linear liquid flux exchanged between the film and the meniscus. However, gravity affects the meniscus size, and consequently the magnitude of the flux. Increasing rotation speed or effective gravity reduces the meniscus radius, thereby enhancing the flux and accelerating film drainage. The scaling follows an exponent of $3/2$ with rotation speed (or equivalently $3/4$ with effective gravity), consistent with theoretical predictions. Geometric effects associated with the frame shape may alter this scaling exponent \cite{vigna_flowing_2025}. A crossover could occur if the meniscus radius $r_{\rm m}$ becomes comparable to the film thickness $h$—a situation expected only at very high rotation speeds ($\omega \approx 3600$~rad·s$^{-1}$, or $\tilde{g} \approx 200$), well beyond the current experimental stability limits. 
%\textcolor{blue}{\sout{Finally, note that the long-time scaling for marginal regeneration–driven thinning, $h_R(t) \propto t^{-2/3}$, decreases more rapidly than the $t^{-1/2}$ scaling expected for thinning by internal flow (Eq.~\eqref{eq:poiseuille}), explaining why marginal regeneration remains the dominant mechanism even at very long times.}}

Future work should focus on a quantitative characterization of the TFE detachment from the meniscus, TFE diameter~$\ell_{\mathrm{TFE}}$ in the film and the detailed, intermittent dynamics of their motion. Such insights would further clarify the mechanisms governing film erosion and the transition between the observed drainage Regimes under extreme effective gravity.

\subsection{Acknowledgments}
This work was supported by the National Research Agency (ANR-20-CE30-0019). The authors are grateful to Isabelle Cantat and Emmanuelle Rio for fruitful discussions.

\bibliography{Monier_bib}% Produces the bibliography via BibTeX.

\end{document}